\documentstyle[aps,preprint]{revtex}

\begin{document}
\author{Remo Garattini}
\address{M\'{e}canique et Gravitation, Universit\'{e} de Mons-Hainaut,\\
Facult\'e des Sciences, 15 Avenue Maistriau, \\
B-7000 Mons, Belgium \\
and\\
Facolt\`a di Ingegneria, Universit\`a degli Studi di Bergamo,\\
Viale Marconi, 5, 24044 Dalmine (Bergamo) Italy\\
e-mail: Garattini@mi.infn.it}
\title{Spacetime foam, Casimir energy and black hole pair creation}
\date{\today}
\maketitle

\begin{abstract}
We conjecture that the neutral black hole pair production is related to the
vacuum fluctuation of pure gravity via the Casimir-like energy. Implications
on the foam-like structure of spacetime are discussed.
\end{abstract}

\section{ Introduction}

It was J. A. Wheeler who first conjectured that spacetime could be subjected
to topology fluctuation at the Planck scale\cite{Wheeler}. This means that
spacetime undergoes a deep and rapid transformation in its structure. This
changing spacetime is best known as ``{\it spacetime foam}'', which can be
taken as a model for the quantum gravitational vacuum. Some authors have
investigated the effects of such a {\it foamy} space on the cosmological
constant, one example is the celebrated Coleman mechanism involving wormholes%
\cite{Coleman}. Nevertheless, how to realize such a foam-like space is still
unknown as too is whether this represents the real quantum gravitational
vacuum. For this purpose, we begin to consider the ``simplest'' quantum
process that could approximate the foam structure in absence of matter
fields, that is the black hole pair creation. Different examples are known
on this subject. The first example involves the study of black hole pair
creation in a background magnetic field represented by the Ernst solution%
\cite{Ernst} which asymptotically approaches the Melvin metric\cite{Melvin}.
Another example is the Schwarzschild-deSitter metric (SdS) which
asymptotically approaches the deSitter metric. The extreme version is best
known as the Nariai metric\cite{Nariai}. In this case the background is
represented by the cosmological constant $\Lambda $ acting on the neutral
black hole pair produced, accelerating the components away from each other.
Finally another example is given by the Schwarzschild metric which
asymptotically approaches the flat metric and depends only on the mass
parameter $M$. Metrics of this type are termed asymptotically flat (A.F.).
Another metric which has the property of being A.F. is the
Reissner-Nordstr\"{o}m metric, which depends on two parameters: the mass $M$
and the charge $Q$ of the electromagnetic field. Nevertheless, all the cases
mentioned above introduce an external background field like the magnetic
field or the cosmological constant to produce the pair and accelerate the
components far away. In this letter, we would like to consider the same
process without the contribution of external fields, except gravity itself
and consider the possible implications on the foam-like structure of
spacetime. This choice linked to the vacuum Einstein's equations leads to
the Schwarzschild and the flat metrics, where only the simplest case is
considered, i.e. metrics which are spherically symmetric. Since the A.F.
spacetimes are non-compact a subtraction scheme is needed to recover the
correct equations under the constraint of fixed induced metrics on the
boundary\cite{Frolov,HawHor}.

\section{Quasilocal Energy and Entropy in presence of a Bifurcation Surface}

Although it is not necessary for the forthcoming discussions, let us
consider the maximal analytic extension of the Schwarzschild metric, i.e.,
the Kruskal manifold whose spatial slices $\Sigma $ represent Einstein-Rosen
bridges with wormhole topology $S^2\times R^1$. Following Ref.\cite{Frolov},
the complete manifold ${\cal M}$ can be taken as a model for an eternal
black hole composed of two wedges ${\cal M}_{+}$ and ${\cal M}_{-}$ located
in the right and left sectors of a Kruskal diagram. The hypersurface $\Sigma 
$ is divided in two parts $\Sigma _{+}$ and $\Sigma _{-}$ by a bifurcation
two-surface $S_0$. On $\Sigma $ we can write the gravitational Hamiltonian 
\[
H_p=H-H_0=\frac 1{2\kappa }\int_\Sigma d^3x(N{\cal H+}N^i{\cal H}_i) 
\]
\begin{equation}
+\text{ }\frac 1\kappa \int_{S_{+}}^{}d^2xN\sqrt{\sigma }\left( k-k^0\right)
-\frac 1\kappa \int_{S_{-}}d^2xN\sqrt{\sigma }\left( k-k^0\right) ,
\label{a1}
\end{equation}
where $\kappa =8\pi G$. The Hamiltonian has both volume and boundary
contributions. The volume part involves the Hamiltonian and momentum
constraints 
\[
{\cal H}=\left( 2\kappa \right) G_{ijkl}\pi ^{ij}\pi ^{kl}-\sqrt{^3g}%
R/\left( 2\kappa \right) =0, 
\]
\begin{equation}
{\cal H}_i=-2\pi _{i|j}^j=0,
\end{equation}
where $G_{ijkl}=\left( g_{ik}g_{jl}+g_{il}g_{jk}-g_{ij}g_{kl}\right) /\left(
2\sqrt{g}\right) $ and $R$ denotes the scalar curvature of the surface $%
\Sigma $. The volume part of the Hamiltonian $\left( \ref{a1}\right) $ is
zero when the Hamiltonian and momentum constraints are imposed. However, for
the flat and the Schwarzschild space, constraints are immediately satisfied,
then in this context the total Hamiltonian reduces to 
\begin{equation}
H_p=\frac 1\kappa \int_{S_{+}}^{}d^2xN\sqrt{\sigma }\left( k-k^0\right) -%
\frac 1\kappa \int_{S_{-}}d^2xN\sqrt{\sigma }\left( k-k^0\right) .
\label{a1a}
\end{equation}
Quasilocal energy is defined as the value of the Hamiltonian that generates
unit time translations orthogonal to the two-dimensional boundaries, i.e. 
\[
E_{tot}=E_{+}-E_{-}, 
\]
\[
E_{+}=\frac 1\kappa \int_{S_{+}}^{}d^2x\sqrt{\sigma }\left( k-k^0\right) 
\]
\begin{equation}
E_{-}=-\frac 1\kappa \int_{S_{-}}d^2x\sqrt{\sigma }\left( k-k^0\right) .
\end{equation}
where $\left| N\right| =1$ at both $S_{+}$ and $S_{-}$. $E_{tot}$ is the
quasilocal energy of a spacelike hypersurface $\Sigma =\Sigma _{+}\cup
\Sigma _{-}$ bounded by two boundaries $^3S_{+}$ and $^3S_{-}$ located in
the two disconnected regions $M_{+}$ and $M_{-}$ respectively. We have
included the subtraction terms $k^0$ for the energy. $k^0$ represents the
trace of the extrinsic curvature corresponding to embedding in the
two-dimensional boundaries $^2S_{+}$ and $^2S_{-}$ in three-dimensional
Euclidean space. Let us consider the case of the static Einstein-Rosen
bridge whose metric is defined as: 
\begin{equation}
ds^2=-N^2dt^2+g_{yy}dy^2+r^2\left( y\right) d\Omega ^2,  \label{a1b}
\end{equation}
where $N$, $g_{yy}$, and $r$ are functions of the radial coordinate $y$
continuously defined on ${\cal M}$, with $dy=dr/\sqrt{1-\frac{2m}r}$. If we
make the identification $N^2=1-\frac{2m}r$, the line element $\left( \ref
{a1b}\right) $ reduces to the S metric written in another form. The
boundaries $^2S_{+}$ and $^2S_{-}$ are located at coordinate values $y=y_{+}$
and $y=y_{-}$ respectively. The normal to the boundaries is $n^\mu =\left(
h^{yy}\right) ^{\frac 12}\delta _y^\mu $. Since this normal is defined
continuously along $\Sigma $, the value of $k$ depends on the function $r,_y$%
, which is positive for $^2B_{+}$ and negative for $^2B_{-}$. The
application of the quasilocal energy definition gives 
\[
E=E_{+}-E_{-} 
\]
\begin{equation}
=\left( r\left| r,_y\right| \left[ 1-\left( h^{yy}\right) ^{\frac 12}\right]
\right) _{y=y_{+}}-\left( r\left| r,_y\right| \left[ 1-\left( h^{yy}\right)
^{\frac 12}\right] \right) _{y=y-}.
\end{equation}
It is easy to see that $E_{+}$ and $E_{-}$ tend individually to the ${\cal %
ADM}$ mass ${\cal M}$ when the boundaries $^3B_{+}$ and $^3B_{-}$ tend
respectively to right and left spatial infinity. It should be noted that the
total energy is zero for boundary conditions symmetric with respect to the
bifurcation surface, i.e., 
\begin{equation}
E=E_{+}-E_{-}=M+\left( -M\right) =0,  \label{a2}
\end{equation}
where the asymptotic contribution has been considered. The same behaviour
appears in the entropy calculation for the physical system under
examination. Indeed 
\begin{equation}
S_{tot}=S_{+}-S_{-}=\exp \left( \frac{A^{+}}4-\frac{A^{-}}4\right) \simeq
\exp \left( \frac{A_H}4-\frac{A_H}4\right) =\exp \left( 0\right) ,
\label{a3}
\end{equation}
where $A^{+}$ and $A^{-}$ have the same meaning as $E_{+}$ and $E_{-}$. Note
that for both entropy and energy this result is obtained at zero loop. We
can also see eqs. $\left( \ref{a2}\right) $ and $\left( \ref{a3}\right) $
from a different point of view. In fact these eqs. say that flat space can
be thought of as a composition of two pieces: the former, with positive
energy, in the region $\Sigma _{+}$ and the latter, with negative energy, in
the region $\Sigma _{-}$, where the positive and negative concern the
bifurcation surface (hole) which is formed due to a topology change of the
manifold. The most appropriate mechanism to explain this splitting seems to
be a black hole pair creation.

\section{Black Hole Pair Creation}

The formation of neutral black hole pairs with the two holes residing in the
same universe is believed to be a highly suppressed process, at least for $%
\Lambda \gg 1$ in Planck's units\cite{Bousso-Hawking}. The metric which
describes such pair creation is the Nariai metric. When the cosmological
constant is absent the SdS metric is reduced to the Schwarzschild metric
which concerns a single black hole. However, one could regard each single
Schwarzschild black hole in our universe as a mere part of a neutral pair,
with the partner residing in the other universe. In this case the whole
spacetime can be regarded as a black-hole pair formed up by a black hole
with positive mass $M$ in the coordinate system of the observer and an {\it %
anti black-hole} with negative mass $-M$ in the system where the observer is
not present. From the instantonic point of view, one can represent neutral
black hole pairs as instantons with zero total energy. An asymptotic
observer in one universe would interpret each such pair as being formed by
one black hole with positive mass $M$. What such an observer would actually
observe from the pair is only either a black hole with positive energy or a
wormhole mouth opening to the observer's universe, interpreting that the
black hole in the ``{\it other universe}'' has negative mass without
violating the positive-energy theorems\cite{Remo,PGDiaz}. This scenario
gives spacetime a different structure. Indeed it is well known that flat
spacetime cannot spontaneously generate a black hole, otherwise energy
conservation would be violated. In other terms we cannot compare spacetimes
with different asymptotic behaviour\cite{Witten}. The different boundary
conditions reflect on the fact that flat space is not periodic in euclidean
time which means that the temperature is zero. On the other hand a black
hole with an imaginary time necessitates periodicity, but this implies a
temperature different from zero. Then, unless flat spacetime has a
temperature $T$ equal to the black hole temperature, there is no chance for
a transition from flat to curved spacetime. This transition is a decay from
the false vacuum to the true one\cite{Coleman1,Perry,Gross,Mazur}. However,
taking account a pair of neutral black holes living in different universes,
there is no decay and more important no temperature is necessary to change
from flat to curved space. This could be related with a vacuum fluctuation
of the metric which can be measured by the Casimir energy.

\section{Casimir Energy}

One can in general formally define the Casimir energy as follows 
\begin{equation}
E_{Casimir}\left[ \partial {\cal M}\right] =E_0\left[ \partial {\cal M}%
\right] -E_0\left[ 0\right] ,
\end{equation}
where $E_0$ is the zero-point energy and $\partial {\cal M}$ is a boundary.
For zero temperature, the idea underlying the Casimir effect is to compare
vacuum energies in two physical distinct configurations. We can recognize
that the expression which defines quasilocal energy is formally of the
Casimir type. Indeed, the subtraction procedure present in eq.$\left( \ref
{a1a}\right) $ describes an energy difference between two distinct
situations with the same boundary conditions. However, while the expression
contained in eq.$\left( \ref{a1a}\right) $ is only classical, the Casimir
energy term has a quantum nature. One way to escape from this disagreeable
situation is the induced gravity point of view discussed in Ref.\cite{B.L.}
and refs. therein. However, in those papers the subtraction procedure in the
energy term is generated by the zero point quantum fluctuations of matter
fields. Nevertheless, we are working in the context of pure gravity,
therefore quasilocal energy has to be interpreted as the zero loop or tree
level approximation to the true Casimir energy. To this end it is useful to
consider a generalized subtraction procedure extended to the volume term up
to the quadratic order. This corresponds to the semiclassical approximation
of quasilocal energy. What are the possible effects on the foam-like
scenario? Suppose we enlarge this process from one pair to a large but fixed
number of such pairs, say $N$. What we obtain is a multiply connected
spacetime with $N$ holes inside the manifold, each of them acting as a
single bifurcation surface with the sole condition of having symmetry with
respect to the bifurcation surface even at finite distance. Let us suppose
the interaction between the holes can be neglected, i.e., let us suppose
that the total energy contribution is realized with a coherent summation
process. This is equivalent to saying that the wave functional support
(here, the semiclassical WDW functional) has a finite size depending only on
the number of the holes inside the spacetime. It is clear that the number of
such holes cannot be arbitrary, but is to be related with a minimum size of
Planck's order. However this amazing mechanism is still to be verified. One
possibility is the computation of 
\[
\Gamma =\frac{P_{N-holes}}{P_{flat}}\simeq \frac{P_{foam}}{P_{flat}},
\]
{\it via} the semiclassical WDW functional. Another way is to be found in
the generalized subtraction procedure, that is, in the evaluation of the
Casimir energy. Nevertheless we obtain, in both cases, a semiclassical
result from which we can extract hints on the possibility of obtaining a
foam-like scenario.

\section{Acknowledgments}

I wish to thank Dott J. Pearson for her invaluable linguistic advice.

\end{document}